# XMM-Newton Observations of NGC 507: Super-solar Metal Abundances in the Hot ISM

**Dong-Woo Kim and Giuseppina Fabbiano**

Smithsonian Astrophysical Observatory,
60 Garden Street, Cambridge, MA 02138

(June 9, 2004)

Abstract

We present the results of the X-ray XMM-Newton observations of NGC 507, a dominant elliptical galaxy in a small group of galaxies, and report 'super-solar' metal abundances of both *Fe* and α-elements in the hot ISM of this galaxy. These results are robust, in that we considered all possible systematic effects in our analysis. We find $Z_{Fe}$ = 2-3 times solar inside the $D_{25}$ ellipse of NGC 507. This is the *highest* $Z_{Fe}$ reported so far for the hot halo of an elliptical galaxy; this high Iron abundance is fully consistent with the predictions of stellar evolution models, which include the yield of both type II and Ia supernovae. Our analysis shows that abundance measurements are critically dependent on the selection of the proper emission model. The spatially resolved, high quality XMM spectra provide enough statistics to formally require at least three emission components in each of 4 circum-nuclear concentric shells (within r < 5 arcmin or 100 kpc): two soft thermal components indicating a range of temperatures in the hot ISM, plus a harder component, consistent with the integrated output of low mass X-ray binaries (LMXBs) in NGC 507. The two-component (thermal + LMXB) model customarily used in past studies yields a much lower $Z_{Fe}$, consistent with previous reports of sub-solar metal abundances. This model, however, gives a significantly worse fit to the data (F-test probability < 0.0001). The abundance of α-elements (most accurately determined by *Si*) is also found to be super-solar. The α-elements to *Fe* abundance ratio is close to the solar ratio, suggesting that ~70% of the Iron mass in the hot ISM was originated from SNe Type Ia. The α-element to *Fe* abundance ratio remains constant out to at least 100 kpc, indicating that SNe Type II and Ia ejecta are well mixed in a scale much larger than the extent of the stellar body.

1. Introduction

Heavy elements in the hot halos of early-type galaxies are the relic of stellar evolution. Determining their abundance is key to our understanding of these galaxies. In particular, since abundances are related to the supernova yield, they can constrain both the supernova rate (Type Ia and II) and the initial mass function (IMF) of the stellar population (e.g., Renzini et al. 1993). Moreover, these measurements are also important for constraining the evolution of the hot ISM in terms of the energy input from supernovae, which may result in the onset of galactic winds. Yet these measurements are difficult and the results have been controversial (see Fabbiano 1995).

Fits of the ROSAT and ASCA data with single temperature thermal spectra (and an additional hard component to account for LMXBs) suggested a hot ISM almost totally devoid of metals (mostly *Fe*) in early-type galaxies (e.g., Awaki et al. 1994; Loewenstein et al. 1994; Davis and White 1996). These results are incompatible with the few-times solar *Fe* abundances predicted by stellar evolution models (e.g., Arimoto et al. 1997). Fitting the X-ray spectra with more complex emission models allowed a higher metal content (e.g., Trinchieri et al. 1994; Kim et al., 1996; Buote and Fabian, 1998; Matsushita et al 2000), but complex models were often not statistically required. While Kim et al. (1996) were the first to be able to reject on statistical grounds the simple model in the case of the ASCA data of NC 4382, an X-ray faint S0 galaxy, they still could not constrain $Z_{Fe}$, because of the limited statistics of these data.

The higher quality Chandra and XMM-Newton data are now showing that the extremely low, sub-solar, Iron abundances suggested by the ROSAT and ASCA analyses can be statistically rejected. In particular, Kim and Fabbiano (2003), by subtracting a population of ~80 discrete sources from the image, excluded sub-solar metal abundances in the hot ISM of NGC 1316. Applying a two-temperature model of the hot gas to XMM-Newton spectra, Buote (2002) reported the first convincing measurement of super-solar metal abundances ($Z_{Fe}$ = 1.5-2 solar) in the central region of NGC 1399. The central Iron abundances which are slightly higher than (or close to) the solar value have been reported for the Virgo cD galaxy, M87 (Gastaldello and Molendi 2002), and nearby galaxy groups, NGC 5044 (Buote et al. 2003) and MKW4 (O'Sullivan et al. 2003). However, the measured abundances are still not as high as the model predictions, ~ a few times solar. Furthermore, there are still some very low abundance reports: for example, O'Sullivan and Ponman (2003) report Z < 0.1 solar in three X-ray faint early type galaxies.

To constrain the heavy element abundance with good quality data of very high statistical significance, we have performed a deep XMM-Newton observation of NGC 507, one of the X-ray brightest early type galaxies in the *Einstein* galaxy sample ($F_X$ ~$10^{-11}$ erg sec$^{-1}$ cm$^{-2}$; Fabbiano, Kim and Trinchieri 1992). After the *Einstein Observatory* discovery of X-ray emission, NGC 507 has been extensively studied in X-rays with the ROSAT PSPC (Kim and Fabbiano 1995; hereafter **KF95**), ROSAT HRI (Paolillo et al. 2003), ASCA (Matsumoto et al. 1997) and Chandra (Forman et al. 2002; Kraft et al. 2003). The ROSAT PSPC observation of NGC 507 revealed a cooler central region, i.e., a positive temperature gradient (**KF95**), typical of bright X-ray ellipticals or small groups of galaxies (e.g., Trinchieri et al. 1997). The values of metal abundances in the hot ISM of NGC 507, reported in the literature, range from 0.2 to near solar (Matsumoto et al. 1997; **KF95**; Buote and Fabian 1998; Paolillo et al. 2003; Kraft et al. 2003).

This paper is organized as follows. In section 2, we describe the XMM-Newton observations and the data reduction. In section 3, we describe the extraction and spectral fitting of X-ray spectra, considering various effects to assess possible systematic errors; these include background subtraction, emission models, atomic emission codes, de-projection of the data, different ways of grouping heavy elements, and fixing/varying $N_H$. In section 4, we present our results on the abundances of *Fe* and α-elements (*Si*, *S*, *Mg* and *O*). In section 5, we discuss the implications of our results in terms of the evolution of the hot ISM and its relation to SNe type Ia and II. Finally, we summarize our conclusions in section 6.

Through this paper, we adopt a distance $D$ = 70 Mpc, based on the heliocentric velocity of 4934 km sec$^{-1}$, or z = 0.016 (Huchra et al. 1999) and $H$ = 70 km sec$^{-1}$ Mpc$^{-1}$. At the adopted distance, 1 arcmin corresponds to 20.4 kpc, and the photometric diameter of NGC 507 $D_{25}$ = 3.1 arcmin corresponds to 63 kpc.

2. XMM-Newton observations

NGC 507 was observed for 40 ksec on Jan. 15, 2001, with XMM-Newton MOS and PN (obsid=0080540101). We use SAS version 5.3 to reduce the data and follow the prescriptions in Snowden et al. (2002). We apply flag = 0 for all instruments and additionally pattern ≤ 4 for PN to exclude bad quality data. No significant background flare is seen during this observation. The effective exposure time is 34.1 ksec for each MOS and 26.6 ksec for PN. We also use CIAO version 3.0 and XSPEC version 11.2 for further analysis.

Figure 1 shows the XMM-Newton (MOS1 + MOS2) X-ray image in the broad band (0.3-8 keV). Also marked in Figure 1 are the $D_{25}$ ellipse of NGC 507 and regions where background spectra are extracted (see section 3). The extended X-ray emission from the hot ISM is seen out to r=10′ (this is consistent with previous studies; see **KF95**; Paolillo et al. 2003). Figure 2 shows the true color image (smoothed with gaussian σ=7.5″) with red for the soft band in 0.3-0.9 keV, green for the medium band in 0.9-2.5 keV and blue for the hard band in 2.5-8.0 keV. This image shows that the emission from the central ~1 arcmin is softer (yellow), as suggested by the ROSAT PSPC data (**KF95)**. Significant sub-structures are visible in this central area in the Chandra image (Forman et al. 2002; Kraft et al. 2003).

Clearly seen in Figure 1 and 2 is also a large number of apparently point-like sources at the periphery of the extended hot ISM emission. The number of these sources is in excess of that expected from background serendipitous sources. These sources were first discovered in the ROSAT PSPC observations (**KF95**). We will present the results of point source properties and spatial analysis of the hot ISM in a future paper. We note here that because these peripheral sources are placed at the XMM-Newton aim point, we could obtain reliable background spectra, determined locally at off-axis distances similar to the sources (see section 3).

3. Spectral analysis

We extract spectra for each instrument (MOS1, MOS2 and PN) from several circular annuli, using **xmmselect**, available in **sas** v5.3. The annuli (with inner-outer radius = 0-1′, 1-2′, 2-3′, 3-5′, 5-7′ and 7-10′) are selected to (a) be at least 1′ wide, so to include a few instrumental beam sizes, and (b) yield at least 5000 counts from each individual instrument after background subtraction, to ensure good statistics. Each spectrum extracted from these regions is then binned to have at least 25 counts in order to perform a chi squared fit. Since we expect very little X-ray emission from the hot ISM (with kT~1 or less) at E > 5 keV, where instead there is strong background emission, we limit spectral fitting to the energy range of 0.3 – 5 keV. We note that increasing the upper limit does not improve statistics, nor change our results. The lower limit is set to avoid the Carbon edge at ~0.28 keV and to remove the contamination from the low pulse-height events (e.g., Snowden et al. 2002). For each spectral extraction annulus, we determined redistribution matrix files (**rmf**) and auxiliary response files (**arf**) with the tools **rmfgen/arfgen** available in **sas** v5.3. Comparing **arf**s for different off-axis distances, we find that telescope vignetting is almost independent of energy at E < 5 keV. Moreover the effective areas of different instruments are consistent at E < 5 keV, but slightly inconsistent at higher energies.

We did not apply any artificial correction factor for different instruments to compensate for systematic cross-calibration uncertainties. Instead, we fitted both individual spectra obtained from MOS1, MOS2 and PN and the joint spectra from MOS1 + MOS2 and MOS1 + MOS2 + PN and present all five sets of results to indicate the possible range of parameters. In general, while the best-fit values of the heavy element abundances may differ, the results are consistent within the 90% error (see section 4.2). Best-fit temperatures (and radial variations) are almost identical (see section 4.1). While the goodness of the fit for each instrument is comparable with each other, PN tends to yields a higher $Z_{Fe}$ than MOS and joint fitting of all 3 instruments gives average best-fit parameters.

In Table 1, we compare the goodness of the spectral fits performed with various combinations of options. In the following sections (3.1 through 3.4) we discuss in detail the effects of systematic uncertainties on the data extraction and fitting.

The baseline case (Fit1 in Table 1) consists of the five different instrument combinations described above. The observed spectra extracted from the first four annuli (0-1′, 1-2′, 2-3′ and 3-5′) are background-subtracted using the local background (see section 3.1) and are fitted simultaneously with projected three dimensional models (section 3.3), where each model representing the emission from a three dimensional shell consists of three spectral components (section 3.2). The heavy elements are constrained to vary together (section 3.4), but the overall amount of elements is fitted independently in each shell. $N_H$ is fixed at the Galactic value (section 3.4). The fit is statistically acceptable with $\chi^2_{red}$ close to 1 (always < 1.2) with 500-2500 degrees of freedom. Note that $\chi^2$ statistics are determined for the entire set of annuli. Also listed in Table 1 are the best-fit $Z_{Fe}$ and the temperature of the soft thermal component (and their 90% acceptable ranges) in the innermost two regions (0-1′ and 1-2′); these are the key quantities we want to measure in this study (see section 4.1 and 4.2). The full range of radial variations of Z and kT are listed in Table 2. Throughout this paper, we quote errors determined at the 90% confidence.

The remaining results (Fit2 – Fit14 in Table 1) are obtained by adopting different data extraction or model fitting assumptions. These include (a) using three different sets of background spectra (from the same observations or from blank-field data), (b) different emission models (2-components and 3-components), (c) different plasma emission codes (mekal and apec), (d) with and without de-projection (2-D and 3-D), (e) $N_H$ fixing at the Galactic value or free to vary, (f) different ways of grouping elements to vary together (Z1 – Z5), and (g) using the data from the fifth (5-7′) and sixth (7-10′) annulus, where the background subtraction is more uncertain (see section 3.3). We note that if not treated correctly, systematic errors could be larger than statistical errors. We will discuss each of them in detail.

For the solar abundance, we adopt the element ratios in Grevesse and Sauval (1998). The new meteoric value for *Fe* is lower (by a factor of 1.48) than the commonly used value in Anders and Grevesse (1989). This change makes $Z_{Fe}$ effectively increase by 50% from those previously determined (e.g., **KF95**), even with no other change.

3.1 Background Spectra

For extended sources, the accurate determination of the field background counts is often a non-trivial task. To determine the effect of the uncertainties in the choice of field background, we have extracted three different sets of background spectra for each instrument. The first set (**BL** in Table 1) is extracted from a circular region close to the aim point, but 7-12′ away from the center of NGC 507 (discrete sources are excluded). In Figure 1, the red circle near the center of the field of view indicates the region of the local background. Because this region is at an off-axis distance comparable to that of our source location, we do not apply corrections for telescope vignetting. The second set (**BE** in Table 1) is extracted from the edge of the detector, 10-19′ away from the center of NGC 507. The three blue circles located in the left and at the bottom of Figure 1 indicate the second background region. To these data we apply a vignetting correction based on the effective area ratios given by the **arf** at E < 5keV. The third set (**BB** in Table 1) is derived from the blank field background data obtained from the XMM-Newton Science Operations Centre (Lumb et al. 2002). In this case, the background spectra are extracted from the same annuli as the source spectra. Because of the temporal and spatial variation of the background count rate (see Lumb et al. 2002), we rescaled the blank field background spectra to our data by comparing the counts extracted from the same location (the red circle in Figure 2) in the energy range of 5-10 keV.

The results of the spectral fits using the three different background spectra are compared in Figure 3 for the MOS1 data (the results from other instruments are similar). The X-ray spectra (and the best fit models) extracted from r < 1′, r = 1-2′, 2-3′ and 3-5′ are shown from top to bottom. Except for the first spectrum, they are vertically displaced for visibility. Strong emission features are also marked. The local background subtraction (**BL**) always gives significantly better fit than the other two (compare Fit1 – Fit3 in Table 1). The spectra obtained with local background subtraction (Figure 3a) work well throughout the whole energy range, resulting in $\chi^2_{red}$ close to 1, and no localized deviations near strong emission features. Instead, using background spectra taken at the edge of the field (**BE**; Figure 3b) or from the blank field data (**BB**; Figure 3c) produces poor fits with $\chi^2_{red}$ of 1.4-1.8 and 1.2-1.3. The F-test indicates that the fit of the BL-subtracted spectra results in a significantly improved $\chi^2$ (a probability of < 0.0007 and 0.05-0.1 for **BE** and **BB**, respectively). The significant deviations at ~1.5 keV (and also at ~2 keV and ~2.3 keV) seen in **BB** and **BE** spectra are likely to stem from incorrect subtraction of the Al K fluorescent emission (Lumb et al. 2002). Also visible at high energies (> 4 keV) are over-subtraction (**BE**) or possibly under-subtraction (**BB**) features.

Based on this analysis, we chose to rely on the **BL** background estimate, except for the outer source annuli (section 3.3) where we use **BB** background, because the local background obtained at r = 7-12' may be contaminated by the X-ray emission of the extended hot ISM (see section 3.3). We note that while the relative fluxes particularly at the outer shell differ, the abundance and temperature are consistent within the error between **BL** and **BB** (compare Fit1 and Fit3 in Table 1).

3.2 Emission Models

The X-ray emission from early type galaxies can consist of many different emission components: hard X-ray emission from low-mass X-ray binaries (LMXB), soft thermal emission from the hot ISM and possibly a power-law component from a low luminosity AGN, not to mention some additional hard emission from incompletely subtracted background sources. In NGC 507 the nuclear emission is not a major component (this will be discussed in paper II). The LMXBs are instead likely to contribute substantially to the hard emission, although we do not see them directly because typical LMXB luminosities are below the detection threshold of these observations (see Section 5). The hot ISM of early type galaxies is likely to consist of multi-temperature gas. For NGC 507, the ROSAT data suggested that the average temperature increases with increasing galactocentric distance (**KF95**). We cannot exclude multi-temperature, possibly inhomogeneous, hot gas (see later in sections 4.1 and 5). Therefore, at least three emission components (two thermal components to give a measure of the range of temperatures in the hot ISM, plus a hard component to model the LMXB emission) are needed to fit the spectra from each annulus. This is a 'minimum' component model, and the real emission could be more complex.

We use a Mewe-Kaastra-Liedahl (mekal) themal plasma model for the thermal emission. We let the temperature of the softer thermal component vary and fix that of the harder thermal component at 1.4 keV, which is the average ISM temperature suggested by the ROSAT data at large galactocentric distances (**KF95**; and see section 4.1). We note that even if the temperature of the harder thermal component is set to vary, it is consistent with being 1.4 keV within the error and our results (including the measured abundance and soft component parameters) remain almost the same. We have also applied an APEC plasma emission code (http://cxc.harvard.edu/atomdb/ source_apec.html) and listed the results in Table 1 (Fit5). The results are similar to those of the baseline case (Fit1). Because LMXBs appear to exhibit uniform spectral properties, we adopt a $kT = 7$ keV Bremsstrahlung component to model their spectrum (e.g., Kim and Fabbiano 2003; Irwin et al. 2003), although the temperature is not critical in our results as long as $kT \gtrsim 3$ keV.

The need of a three-component (**3C**) model is demonstrated by a comparison with two-component (**2C**) fits in Figure 4a and 4b. We find a significant improvement with the three-component model, clearly indicating the presence of thermally complex ISM (also compare Fit1 and Fit4 in Table 1). F-tests for **3C** over **2C** indicate very low probabilities to exceed the given F-statistic ($< 10^{-4}$). The localized deviation near the *Fe* peak (~1 keV) is clearly seen in the two-component fit (Figure 4b). This kind of deviation is reminiscent of the results of spectral fits with previous X-ray missions, which resulted in an extremely low Z with a small error (e.g., Matsushita et al. 1994; Matsumoto et al. 1997). We reproduce the same trend that the best-fit $Z_{Fe}$ in the two-component fit is significantly lower than that of the three-component fit (see section 4.2).

It is known that some atomic data may be uncertain, particularly for Fe XVII lines (e.g., Xu et al. 2003). To test the effect of these uncertainties on our results, we have re-fitted the XMM spectra after (1) excluding the energy ranges (0.7-0.75 keV and 0.8-0.85 keV) of Fe XVII lines (15.01, 16.78, 17.951, 17.096Å) and (2) adjusting the line ratios (increasing by ~50% of 16.78, 17.951, 17.096Å lines) based on Capella data (Canizares et al. 2000). In both tests, our results remain the same. This is because Fe XVII lines are not very strong in NGC 507 due to the ISM temperature higher than 0.6 keV (0.8-1.4 keV in the center; see Figure 3 where the Fe peak is at ~1 keV, instead of 0.7-0.8 keV) and because the CCD spectra are not sensitive to over/under-estimated individual lines. This is also confirmed by the identical results obtained with different emission codes (MEKAL vs APEC). Xu et al. (2002) point out that some strong lines (e.g., Fe XVII 15.01Å) may be optically thick near the center. We did not correct for resonance scattering in our analysis. However, we expect the resonance scattering will not change our results as indicated in the above test (1); this effect, if added, will only increase the measured abundance.

3.3 De-projection

Since the ISM is not isothermal, and there is a radial temperature gradient, we need to 'de-project' each spectrum obtained from a two-dimensional annulus, assuming spherical symmetry. To this end we used **project,** available in **xspec** v11.2, where three-dimensional models (representing spectra from three-dimensional shells) are projected into a plane and simultaneously fit to a set of observed spectra extracted from multiple annuli. We compare the results of three-dimension (**3D**) vs. two-dimension (**2D**) fitting in Figure 4a and 4c (or compare Fit1 and Fit6 in Table 1, for the **3C** model). Similarly, for two-component models, **3D** and **2D** results are compared in Figure 4b and 4d (or compare Fit4 and Fit7 in Table 1). Although the difference between **3D** and **2D** fits is less significant (and with similar reduced $\chi^2$) than that between **3C** and **2C** models, the **2D** fits produce a lower $Z_{Fe}$ than the **3D** in all cases. We note that the combination of a two-component model and **2D** fitting yields the lowest $Z_{Fe}$: this is the typical approach found in published work (see section 5).

To most accurately determine abundances in the hot ISM, our baseline case (Fit1 in Table 1) used the spectral data up to r=5′ or ~100 kpc and the local background (**BL**). As described in section 3.1, **BL** provides the best $\chi^2$ statistics. However, the diffuse X-ray emission extending to 10′ (e.g., **KF95**) may affect our results in the inner region. To quantify the projection effect of the outer region (from 5′ to 10′), we repeat the de-projection, by adding the outer two shells 5-7′ and 7-10′ (Fit13-14 in Table 1). As the local background may contain emission from the extended sources, we apply the background spectra obtained from the blank field data (**BB**) and compare results with different outer shells, but with the same background data (i.e., Fit3 and Fit13-14 in Table 1). The results in the inner four shells (< 100 kpc) are consistent within the errors. In particular, the metal abundances remain unchanged (see also Table 2).

Since there have been reports of central minima (r < 0.5′) of metal abundances, e.g., in M87 (Gastaldello and Molendi 2002) and in NGC 5044 (Buote et al. 2003), we have also divided the central bin (r < 1′) in two parts (r < 0.5′ and r = 0.5 - 1′), to check for small-scale variation in the central region. We do not find any significant difference inside r < 1′ of NGC 507 (this is not surprising given the XMM PSF size; Ehle et al. 2003): $Z_{Fe}$ increases slightly toward the center (MOS) or remains constant (PN).

3.4 Linking element abundances

Given the limited statistics and the various systematic effects (e.g., blending of emission features in the low-resolution CCD spectra and the presence of a strong Al K fluorescent line), it is not always possible, nor optimal, to vary all the heavy elements independently. We have therefore linked different group of elements to vary together. First (**Z1**), all elements vary together with *Fe* and their relative abundances are fixed at the solar ratio given by Grevesse and Sauval (1998). Second (**Z2**), *Si* and *S* (the two most prominent elements among α-elements) vary together and the rest elements are tied to *Fe*. Third (**Z3**), elements lighter than *Ar* are tied to *Si* and those heavier than *Ca* to *Fe*, to distinguish α-elements and those mostly produced by SN Type Ia. Fourth (**Z4**), *Fe* and *Si* are fixed at the average best-fit values in **Z1-Z3** and *S*, *Mg* and *O* vary independently while the other elements are tied as in **Z3**. Fifth (**Z5**), *Fe*, *Si*, *S*, *Mg* and *O* vary independently and the other elements are tied as in **Z3**. In general, while the goodness of the fit is almost the same regardless of different grouping, the best-fit abundances vary somewhat (see Section 4.4).

We tried to independently vary the abundances of two thermal components (in 3C models) to test the abundance inhomogeneity, where the metal rich gas could cool more rapidly. However, with our statistics, we could not see any significant difference. The accurate abundance measurement of individual components will require much higher S/N and higher resolution data.

3.5 $N_H$

Finally, we either fix $N_H$ to be the Galactic value (5 x $10^{20}$ cm$^{-2}$) or let it vary freely (Fit8 in Table 1). The best-fit $N_H$ is consistent with the Galactic value at ~3σ. We find that $N_H$ often goes together with the amount of the hard component and affects $Z_{Fe}$, in the sense that a larger $N_H$ and a larger hard-component effectively reduce $Z_{Fe}$. We will discuss this effect in section 5.

4. Results

4.1. Temperatures of multi-component emission models

Our results are consistent with the overall positive temperature gradient, suggested by the ROSAT PSPC data (**KF95**). In addition, we find that the hot ISM within each three-dimensional shell is not isothermal and that a range of kT is required. In Table 2, we list the spectral parameters determined in each shell, obtained by jointly fitting the spectra from all the instruments (**MOSPN** in Table 1). We compare the radial variation of the relative contributions from different emission components in Figure 5a and 5b. We plot the results of the baseline case (Fit 1 in Table 1), note that other 3C models yield similar distributions of temperatures and flux ratios. The temperature of the 1$^{st}$ soft component (blue circles in Figure 5a) ranges from 0.6 to 0.8 keV (see Figure 5a). The X-ray flux ratio of the two thermal components measures the relative importance of cooler (0.6-0.8 keV; blue circles in Figure 5b) and hotter (1.4 keV; open red triangle in Figure 5b) emission. In the innermost shell (r < 20 kpc), the contributions from the two components are roughly equal, while in the second shell (r = 20 – 40 kpc), there is a smaller amount of cooler component than in the central bin, with a ratio of 1:3. In the outer shells (r > 40 kpc), the ratio drops to 1:12, and the gas at ~1.4 keV dominates the X-ray emission. Note that the cooler component is still required in the outer shells, even though it is relatively small.

In Table 2 (at the bottom) and Figure 5a (marked by x), we also show the best-fit gas temperatures obtained with two-component models (Fit4 and Fit7 in Table 1). Although the fits are not acceptable, we can consider the resulting temperatures as the emission-averaged temperatures of the hot ISM. This emission-averaged temperature is ~1 keV near the center and increases to ~1.4 keV at the outskirts, consistent with the measurements in **KF95**. We also tried cooling flow models (vmkcflow in xspec). In this case, the low *kT* is ~0.6 keV and the high *kT* is ~1.4 keV near the center, but the fit is poor ($\chi^2_{red}$ ~ 1.5). As reported for other early type galaxies and clusters (Buote et al. 2003; Peterson et al. 2001), we detect no thermal emission from gas with *kT* < 0.6 keV, for any choice of emission model.

The third (hard LMXB) component (marked by squares in Figure 5b) accounts for ~7% of the total X-ray emission in the first 2 shell (i.e., within the optical galaxy, r < 40 kpc), which is consistent with the expected amount from LMXBs (see section 5). Its amount is undetermined (or possibly absent) in the third shell (r = 40 – 60 kpc). Instead, a non-negligible amount of hard component is required in the fourth shell (r = 60-100 kpc). This hard emission may originate from background AGNs, or may be due to sources similar to the sources detected at larger distances, in excess of the deep survey background predictions (this will be further discussed in paper II.)

In summary, (1) the emission-averaged temperature of the hot ISM increases with increasing distance from the center (from ~1.0 keV to ~1.4 keV); (2) the hot ISM is not locally isothermal: at least two components are needed to fit the data, with the cooler one (*kT* ~ 0.6-0.8 keV) contributing proportionally more at smaller radii; (3) at least 3 emission components, including a hard component probably from a population of sources unrelated to the hot ISM – e.g., LMXBs –, are required.

4.2. Iron abundance

The radial distribution of the *Fe* abundance is shown in Figure 6a (see also Table 2). The best-fit $Z_{Fe}$ is ~3 times solar in the center (r < 1′ or < 20 kpc), ~2 times solar at r = 20-40 kpc and similar to (or slightly lower than) solar at r > 40 kpc. The acceptable range of $Z_{Fe}$, given by the statistical error at 90% confidence is roughly ±1 solar in fitting the spectra from individual instruments, or ±½ solar in the joint fits of the spectra from all the instruments. The maximum $Z_{Fe}$ could be as high as 4 times solar inside the $D_{25}$ ellipse of the galaxy.

The presence of super-solar $Z_{Fe}$ in NGC 507, and its negative radial gradient, are robust results, as shown by the comparison of the fits obtained with the different approaches used in this study. Varying $N_H$ slightly reduces $Z_{Fe}$ (Fit8, see also section 5). Different methods of grouping the heavy elements (Fit1 and Fit9-12 in Table 1) result in similar values of $Z_{Fe}$. Fit12 (**Z5** where Fe, Si, S, Mg, and O vary independently) appears to produce the lowest $Z_{Fe}$, but it is still consistent with other results within the acceptable range. In this case, $Z_{Fe}$ in the second shell remains similar to that in the center (i.e., a shallow gradient). The super solar $Z_{Fe}$ is not affected by the uncertainty in the background spectra. Although the fit with different background spectra (**BE** and **BB**) may be poor (see Fit2 and Fit3), $Z_{Fe}$ still ranges between 2-4 times solar in both cases, consistent with **BL**. Also we note that the *Fe* abundance measured in the central 2′ region will be the least affected by background uncertainties. Also the uncertainties of the plasma emission codes do not affect our results (see section 3.2). Considering all these various systematic effects, we conclude that $Z_{Fe}$ is 2-3 times solar within the optical confines of NGC 507.

We note that the lowest $Z_{Fe}$ value is obtained with the two-component (**2C**) model + **2D** fitting (Fit7 in Table 1; Figure 4d). In this case, the best-fit $Z_{Fe}$ is ~1 solar, which would be ~0.6 solar with Anders and Grevesse (1989) solar ratios, close to the previously reported sub-solar abundance (e.g., Matsushita et al. 2000; see also Kraft et al. 2003). The effect of the adoption of a two-component model on the abundance measurement will be further discussed in section 5. Recent analysis of XMM-Newton RGS data suggested that the metal abundance is sub-solar in NGC 4636 (Xu et al. 2002) and NGC 533 (Peterson et al. 2003). Although the RGS data have high spectral resolution, the analysis of slit-less spectra is complicated and depends heavily on a considerable amount of Monte Carlo simulations and modeling. Future independent confirmation of these results would be desirable.

4.3. Abundance ratios of different heavy elements

Next to *Fe, Si* has the second strongest emission features. These lines are relatively isolated around ~2 keV and hence provide the most reliable measurement of $\alpha$-element abundances. In Table 2 we list the results obtained by different grouping of heavy elements (**Z2** and **Z3**; see section 3.4). Although these results are slightly method-dependent, **Z2** giving a slightly higher $Z_{Si}$ than **Z3**, they are all consistent at the ~$2\sigma$ (or better) significance. On average, the best-fit $Z_{Si}$ ranges from 2 to 3 times solar within the $D_{25}$ ellipse and decreases to ~1 solar outside the optical galaxy, generally following the behavior of the Iron abundance. The ratio of *Si* to *Fe* is therefore close to solar in all radial bins, i.e., there is no radial gradient of [*Si/Fe*] (see Figure 6b). The solar ratio of *Si/Fe* is consistent with previous reports (e.g., Matsushita et al. 2000), although the absolute abundances in these studies ($Z_{Fe}$ or $Z_{Si}$) are generally lower than those we find here. This ratio is also consistent with the [*Si/Fe*] - *kT* relationship of galaxies and clusters (e.g., Fukazawa et al. 1998).

Abundance measurements of the other $\alpha$-elements are not as reliable as *Si*. This is because their emission features are weaker and/or confused, and in some cases (*Mg*, *O*) affected by calibration uncertainties. *S*, *Mg* and *O* are set to vary independently in **Z4** and **Z5** (section 3.4). Although the reduced $\chi^2$ values are almost the same in the two cases (compare Fit 11 and Fit12 in Table 1), the best-fit abundances are higher in **Z4** than **Z5**, but the differences are always within the 90% error.

The emission features of *S* are found next to those of *Si* around E ~ 2.5 keV. They are also relatively isolated, but weaker than those of *Si*. The best-fit $Z_S$ is slightly lower than *Fe* or *Si*. *S/Fe* is ~0.6 and barely consistent with solar in the 90% confidence. No significant radial variation of *S/Fe* is evident.

The *Mg* emission features around E ~ 1.5 keV can be easily identified in thermal gas emission with *kT* < 1 keV. However, *Fe* features start to blend with the *Mg* features in the hotter X-ray plasma with $kT > 1$ keV. Because the hot ISM in NGC 507 has at least two temperatures in the range $kT = 0.6$-$1.4$ keV (see section 4.1), the *Mg* features are somewhat mixed with the *Fe* emission. The *Mg* abundance is also somewhat uncertain because also present at E ~ 1.5 keV is a strong Al K fluorescent line from the camera body (Lumb et al. 2002). The best-fit $Z_{Mg}$ is slightly lower than *Fe* or *Si*. *Mg/Fe* is ~0.8, but consistent to be solar in the 90% confidence. As in the case of Si and S, there is no significant radial variation of *Mg/Fe*.

The emission features of *O* are at E ~ 0.6-0.7 keV (or ~0.5 keV for the colder plasma). They are relatively isolated, but partially blended (at E ~ 0.7 keV) with the *Fe* features in plasmas with $kT = 0.5 - 1$ keV. The best-fit *O/Fe* is lowest (0.3-0.5 solar) among the measured α–elements within the $D_{25}$ ellipse, while consistent to be solar at larger distances. The significance of deviation from the solar ratio is ~4σ at r < 20 kpc and ~3σ at r = 20-40 kpc. Given that the instrument calibration is least accurate at lower energies and that absorption may affect the result in this band, we consider the under-abundant *O* as suggestive, but not conclusive, evidence.

5. Discussion

5.1 Elemental abundances and Supernova yields

As discussed in section 1, the low, often sub-solar, abundance of heavy elements, and particularly Iron, reported in the hot ISM of early type galaxies has been the subject of controversy (see Fabbiano 1995). Iron, in particular, which exhibits the strongest X-ray emission features, is predicted to have 2-5 times solar abundance in the hot ISM (see Arimoto et al. 1997): the *Fe* abundance in the hot ISM is expected to be at least similar to (or higher than) that of the stellar population in elliptical galaxies, where Iron was initially synthesized by the bulk of Type II supernova (SN) explosions and then enriched during the lifetime of the galaxy by Type Ia SNe.

While we had in the past suggested that these apparent low metal abundances are the result of hidden complexity of the X-ray spectra (see discussion in Fabbiano 1995; Kim et al. 1996; see also Buote and Fabian 1998), this conclusion was hard to prove with the then available data. Our XMM-Newton spectra of NGC 507 clearly require a departure from a simple locally isothermal emission model for the hot ISM, resulting in a robust determination of super-solar *Fe* abundances, fully in agreement with the metal enrichment theory (e.g., Arimoto et al. 1997): $Z_{Fe}$ ~ 2-3 times solar and possibly up to ~4 times solar at the center of NGC 507. The measured Iron abundances indicate a negative radial gradient (Figure 6a). Because the stellar density profile is much steeper than the gas density profile (for example, $\rho_* \sim r^{-3}$ while $\rho_{gas} \sim r^{-1.5}$, if $\Sigma_{opt} \sim \Sigma_X \sim r^{-2}$), the metal enrichment by mass-loss and SN ejecta has been more significant near the center than the outskirts.

We also detect emission from *Si*, *S*, *Mg*, and *O*, and measure abundances for these elements. Determining the relative abundance of *Fe* and α-elements is critical for discriminating between the relative importance of SN type II and type Ia in the parent galaxy (e.g., Renzini et al. 1993; Loewenstein et al. 1994). Therefore, these measurements provide important clues for our understanding of the evolution of both stellar component and hot ISM. If heavy elements are mainly synthesized in Type II SNe, the abundance ratio of α-elements to *Fe* is expected to be higher than the solar ratio (e.g., Woosley et al. 1995), while the ratio decreases as increasing contribution from Type Ia SNe (e.g., Iwamoto et al. 1999). In section 4.3, we show that the abundance ratio of *Si* to *Fe* is close to the solar ratio. Note that among α-elements the abundance measurement of *Si* is least uncertain because of its strong, isolated emission features; the theoretical yields of *Si* are also the best determined, with the least amount of scatter between model predictions (e.g., Gibson et al. 1997; Nagataki & Sato 1998). With SN yields taken from Gibson et al. (1997) and converted to the revised solar values given by Grevesse and Sauval (1998), the measured abundance ratio of *Si/Fe* (near solar) indicates that 60-80% of the detected Iron mass is produced in SN Type Ia.

*S* and *Mg* are found to be slightly less abundant (relative to *Fe*) than *Si*, but their abundance ratios are still consistent with the solar ratio within the statistical errors. *O* is the least abundant among the measured α-elements, appearing to have sub-solar abundance (~3σ) within the $D_{25}$ ellipse of the galaxy (out to a radius of ~40 kpc) and possibly increasing to the solar abundance ratio in the outskirts (at radii of 40-100 kpc). Because both *Mg* and *O* are mainly produced by SNe type II, the apparent decrease of the *O/Mg* at the center of NGC 507 is hard to interpret with a simple combination of SNe type II and Ia. Buote et al. (2003) reported a similar trend of low *O* abundance in the center of NGC 5044 (a galaxy similar to NGC 507 in both optical and X-ray properties) and suggested a warm absorber or an unknown physical/instrumental effect. Given the uncertainties discussed in section 4.3, we consider that the sub-solar *O* abundance in the central regions of NGC 507 requires further confirmation.

While we see a significant radial variation in metal abundances in both *Fe* and α-elements, the ratio of *Fe* to α-element abundance appears to be constant, although a mild radial gradient, either positive or negative, may be allowed by the large error bars at large radii (see Figure 5d). This constant ratio suggests that SNe Type II and Ia ejecta are well mixed in a larger scale (~100 kpc) than the optical galaxy (~40 kpc), contrary to the conclusions based on the sub-solar *Fe* abundances estimated with ASCA spectra (e.g., Mushotzky et al. 1994; Matsumoto et al. 1997). These results, which we believe were biased by the assumption of a simple emission model (see Section 3), led to the suggestion of a considerably flatter IMF (initial mass function) and less important SN Ia activity in elliptical galaxies (Loewenstein et al. 1994), than had been assumed on the basis of stellar evolution models for these systems (e.g., Renzini et al. 1993; Arimoto et al. 1997). Our results indicate that these constraints on the IMF and/or a reduced SN Ia rate are not needed.

5.2 The hot ISM (temperature and absorption)

As discussed in section 3.2, the temperature of the hot ISM is not simply a function of the galactocentric radius. In addition to the overall temperature gradient detected in the ROSAT data (**KF95**), the hot ISM appears to consist (at each radius in a three-dimensional distribution) of at least two gaseous components with different temperatures. While the 1.4 keV component dominant at large radii may be representing the virialized hot halo, our result of a multi-phase hot medium suggests that the central region of NGC 507 exhibits complex sub-structures, maybe resulting from dynamical perturbations. These structures include two distinct emission peaks, possibly separated by the nuclear radio jet, seen in the high resolution Chandra image (Forman et al. 2002; Paolillo et al. 2003). Also present are discontinuities of the X-ray surface brightness toward the NE (Kraft et al. 2003) and the SW (in paper II). These features suggest either contact discontinuities or cold fronts, which could be indicative of recent mergers. In cosmological simulations of elliptical galaxies and clusters, a finite range of temperatures at a given radius is often found (e.g., Kawata and Gibson 2003). Therefore, it is clear that an overall radial temperature gradient alone does not reflect the real properties of the hot ISM.

As reported for other early type galaxies and clusters (e.g., Peterson et al. 2001; Buote et al. 2003), we also find no thermal emission from gas cooler than $kT < 0.6$ keV (roughly ½ of the temperature in the ambient gas). Some heating mechanisms (such as AGN feedback or thermal conduction) could compensate the radiative cooling (e.g., Fabian et al. 2003). The lack of cooling below the observed limit is sometimes used to argue against a multi-temperature model. However, this does not rule out temperatures in the observed range, between that in the ambient gas (~1.4 keV) and ~0.6 keV.

The best-fit hydrogen column ($N_H = 6-7 \times 10^{20}$ cm$^{-2}$ in Fit8 of Table 1) is close to the Galactic line-of-sight value ($5 \times 10^{20}$ cm$^{-2}$). Our estimate is consistent with the ROSAT results (**KF95**), but considerably lower than the ASCA reports of $N_H = 1-2 \times 10^{21}$ cm$^{-2}$, which suggested absorption intrinsic to NGC 507 (Matsumoto et al. 1997 and Matsushita, et al. 2000). Neither IRAS FIR (Knapp et al. 1989) nor HI observations (Knapp et al. 1985) of NGC 507, however, indicate significant internal absorption, yielding an upper limit of $M_{HI}$ of $2.7 \times 10^9$ M$_\odot$. Since an intrinsic hydrogen column of a few $\times 10^{20}$ cm$^{-2}$ within the D$_{25}$ ellipse of NGC 507 would correspond to $M_{HI} \sim 10^{10}$ M$_\odot$, we can rule out the presence of internal absorption in NGC 507, unless there is a significant amount of molecular gas (see Arabadjis and Bregman 1999). We note that $N_H$ and the amount of hard component returned by the spectral fits are partially tied, in the sense that a larger $N_H$ tends to go with a smaller hard component. This in turns would affect the model predictions for the thermal continuum at low (E < 0.7 keV) and high energies (E > 2 keV), reducing the required strength of the Fe peak at ~1 keV. This effect could also be partly responsible for ASCA estimates of low abundances.

5.3 The hard spectral component

How much X-ray emission do we expect from the LMXBs in NGC 507? Is this emission consistent with that inferred from our best-fit hard emission component? Chandra observations of giant elliptical galaxies have detected populations of discrete, point-like sources, mostly LMXBs associated with the galaxies (e.g., ~150 in NGC 1399; Angelini et al. 2002). Because of the distance to NGC 507, typical LMXBs cannot be detected in either our XMM-Newton observation, or in the existing 16 ksec Chandra observation. In this Chandra image (obsid=00317) only 3 non-nuclear sources can be barely detected within the $D_{25}$ ellipse (Paolillo et al. 2003; also to be presented in paper II). At a distance of 70 Mpc, typical LMXBs (with $L_X=10^{37} - 10^{38}$ erg sec$^{-1}$) in NGC 507 would result in less than 1 count in the above Chandra observation. Only ultra-luminous X-ray sources (ULX; with $L_X > 10^{39}$ erg s$^{-1}$) could be possibly detected. The total X-ray luminosity of undetected LMXBs can be determined indirectly by using its relationship with the optical luminosity. We use here the relationship determined by Kim and Fabbiano (2004) using a large sample of early type galaxies observed with Chandra, for which these authors derived incompleteness-corrected X-ray luminosity function within the $D_{25}$ ellipse:

$$L_X(\text{LMXB})/L_B = 0.9 \pm 0.5 \times 10^{30} \text{ erg sec}^{-1} / L_{B\odot},$$

where $L_X$ is measured in 0.3-8 keV and $L_B$ is measured in unit of $L_{B\odot}$ adopting $M_{B\odot} = 5.47$ mag. We estimate $L_X(\text{LMXB}) = 1.2 \pm 0.7 \times 10^{41}$ erg sec$^{-1}$ for $B_T^o = 12.19$ mag (taken from RC3), or $F_X(\text{LMXB}) = 2.0 \pm 1.1 \times 10^{-13}$ erg sec$^{-1}$ cm$^{-2}$, which is ~2% of the total X-ray emission within 100 kpc, or ~10% of $L_X$ within D25 (r < 40 kpc). The flux of hard component determined by our spectral fitting in this paper is in excellent agreement with the above estimate (Table 2). The best-fit normalization of the hard component (7 keV Bremsstrahlung) inside r < 2' (or 40 kpc) ranges between 2 to 3 × $10^{-13}$ erg sec$^{-1}$ cm$^{-2}$.

As discussed in section 4.1, a non-negligible amount (a few × $10^{-13}$ erg sec$^{-1}$ cm$^{-2}$) of hard emission is found at radii outside the main stellar body of NGC 507 (r = 60-100 kpc), where we would not expect to find a significant amount of LMXBs. This 'external' hard emission can not be explained with background AGNs: based on the Log(N)-Log(S) relationship determined in ChaMP (Kim et al. 2004), the expected X-ray flux of background sources within the annulus (r = 60-100 kpc) is about 10-20% of the observed hard component. Another possibility is that this 'external' hard emission may be related to the population of sources that we detect in the outer halo of NGC 507, in excess of the expected background sources (see Figure 1). We will discuss these sources in a future paper.

6. Conclusions

We have presented the spatially resolved spectral analysis of the XMM-Newton observations of the halo-dominated X-ray emission of the elliptical galaxy NGC 507. After considering different effects in our spectral fitting, we conclude:

1. While the temperature of the hot ISM increases with increasing galactocentric radius as previously reported, the local ISM is not isothermal. Three-component emission models (two thermal gas components plus additional hard LMXB emission) are needed to model these data.

2. With these models, we find that the *Fe* abundance is super-solar (2-3 times solar) within the stellar body of NGC 507 (a radius of ~40 kpc). The allowed maximum limit is ~4 times solar. This is the highest $Z_{Fe}$ reported for the hot ISM of an early-type galaxy, and it is fully consistent with the abundance predicted by the stellar evolution models. The *Fe* abundance decrease with galactocentric radius to values close to solar outside the optical galaxy, out to r = 100 kpc.

3. The $\alpha$-element abundances (mainly determined by *Si*) are also super-solar and the *Fe* to $\alpha$-element abundance ratio is close to the solar ratio. The *Fe* to *Si* ratio suggests that 60-80% of the *Fe* mass is originated from SN Type Ia.

4. While the *Fe* and $\alpha$-element abundances decrease with increasing radius, their ratio remains solar out to 100 kpc. This, in addition to the near solar $Z_{Fe}$ at large radii, indicates that SNe Type II and Ia ejecta are well mixed throughout the hot ISM.

5. The hot ISM is likely in an inhomogeneous multi-phase state with temperatures ranging from 0.6 keV to 1.4 keV (within a shell at a constant galactocentric distance). However, no cooling below 0.6 keV is identified. This is possibly related to the heating by the AGN as indicated by the radio jet or by thermal conduction.

6. Although we do not detect individual LMXBs in NGC 507, our spectral analysis indicates a hard component of $F_X$ = 2-3 x $10^{-13}$ erg sec$^{-1}$ cm$^{-2}$, which is fully consistent with the expected amount, based on the $L_X$(LMXB) – L(B) relation of early type galaxies.

This work was supported by NASA grant NAG5-9965. We thank Nancy Brickhouse for useful discussions about emission codes and their uncertainties, and the XMM-Newton User Support team and GSFC Guest Observer Facility for their help in data reduction.

Figure Captions

Figure 1. XMM-Newton (unsmoothed) image. The $D_{25}$ ellipse of NGC 507 is marked with a white ellipse. Also marked are regions from where the three sets of background spectra are extracted. The local background is taken from the red circle near the aim point. The second background extraction region is represented by the three large blue circles near the edge of the field of view. These same regions are also used for scaling the counts from blank field data (see text for more details). North is to the top and east is to the left.

Figure 2. XMM-Newton true-color X-ray image of NGC 507. MOS1 and MOS2 images are combined and smoothed with a gaussian $\sigma=7.5''$. Red represents the soft band (0.3-0.9 keV), green the medium band (0.9-2.5 keV) and blue the hard band (2.5-8.0 keV). North is to the top and east is to the left.

Figure 3. Comparison of the results obtained for data extracted with three different sets of background spectra. (a) Source spectra are extracted from 4 concentric annuli and background-subtracted using the local background spectra obtained near the aim point. The data are then fitted together with projected 3-dimensional models with three emission components (see text). All the heavy elements vary together (but independently in different shells) and $N_H$ is fixed at the Galactic value. From top to bottom, shown are the X-ray spectra extracted from $r < 1'$, $r = 1-2'$, $2-3'$ and $3-5'$. Except for the first spectrum, they are vertically displaced for visibility. (b) same as (a) but with background spectra taken from the edge of the field of view and re-scaled using the arf at $E < 5$ keV. (c) same as (a) except the background spectra are taken from the blank field data and re-scaled by the ratio of counts taken from the local background region at $E = 5 - 10$ keV.

Figure 4. Comparison of spectral fitting with 3-component vs. 2-component models and 3-D vs. 2-D. (a) Same as Figure 3-a, i.e., for 3-component models and 3-D fitting, (b) 2-component emission models and 3-D fitting. (c) 3-compnent models and 2-D fitting, and (d) 2-compnent models and 2-D fitting.

Figure 5. (a) Radial distribution of temperatures in the hot ISM. The circles (in blue) and triangles (in red) represent the cooler and hotter components in 3-component models, respectively. The asterisks represent emission-averaged temperatures determined with 2-component models. (b) Radial distribution of relative fluxes of 3 emission components. The circles (in blue) and triangles (in red) represent the cooler and hotter thermal components, respectively, while the squares (in black) the hard LMXB component.

Figure 6. (a) Radial distribution of *Fe* abundances The asterisks (black), circles (red), squares (blue) and cyan (triangle) are determined by **Z1**, **Z2**, **Z3**, and **Z5** (see Table 1). (b) Radial distribution of *Si* to *Fe* abundance ratio. The circles (red), squares (blue) and cyan (triangle) are determined by **Z2**, **Z3**, and **Z5** (see Table 1).

Table 1
Goodness of Spectral Fitting

| Method | red_Ch2 (Ch2 / DoF) | Z(Fe) 0'-1' (0-20 kpc) | Z(Fe) 1'-2' (20-41 kpc) | kT 0'-1' (0-20 kpc) | kT 1'-2' (20-41 kpc) |
|---|---|---|---|---|---|
| <Fit1: baseline> | | | | | |
| BL 3C 3D FNH Z1 MOS1  | 1.05 (  578 /  554 ) | 2.56 (1.7-3.8) | 1.82 (1.2-2.4) | 0.85 (0.81-0.89) | 0.79 (0.73-0.84) |
| BL 3C 3D FNH Z1 MOS2  | 1.16 (  620 /  535 ) | 2.43 (1.5-4.9) | 2.24 (1.5-3.3) | 0.83 (0.78-0.86) | 0.81 (0.75-0.86) |
| BL 3C 3D FNH Z1 MOS12 | 1.17 ( 1296 / 1109 ) | 2.19 (1.5-3.1) | 2.19 (1.6-3.0) | 0.84 (0.81-0.87) | 0.80 (0.77-0.84) |
| BL 3C 3D FNH Z1 PN    | 1.21 ( 1731 / 1428 ) | 3.67 (2.3-5.7) | 2.32 (1.8-3.1) | 0.79 (0.76-0.82) | 0.76 (0.71-0.80) |
| BL 3C 3D FNH Z1 MOSPN | 1.21 ( 3092 / 2557 ) | 2.85 (2.2-3.8) | 2.16 (1.8-2.6) | 0.82 (0.79-0.84) | 0.78 (0.75-0.80) |
| <Fit2 Fit3: different background> | | | | | |
| BE 3C 3D FNH Z1 MOS1  | 1.37 (  758 /  554 ) | 1.99 | 1.61 | | |
| BE 3C 3D FNH Z1 MOS2  | 1.83 (  977 /  535 ) | 1.58 | 1.58 | | |
| BE 3C 3D FNH Z1 MOS12 | 1.68 ( 1862 / 1109 ) | 1.98 | 1.65 | | |
| BE 3C 3D FNH Z1 PN    | 1.53 ( 2184 / 1428 ) | 2.45 | 2.08 | | |
| BE 3C 3D FNH Z1 MOSPN | 1.67 ( 4269 / 2557 ) | 2.00 | 1.79 | | |
| | | | | | |
| BB 3C 3D FNH Z1 MOS1  | 1.19 (  658 /  554 ) | 2.31 (1.7-3.2) | 1.81 (1.2-2.8) | 0.86 (0.82-0.90) | 0.80 (0.72-0.84) |
| BB 3C 3D FNH Z1 MOS2  | 1.31 (  703 /  535 ) | 2.22 (1.4-4.9) | 2.32 (1.5-4.0) | 0.83 (0.79-0.86) | 0.82 (0.76-0.86) |
| BB 3C 3D FNH Z1 MOS12 | 1.29 ( 1430 / 1109 ) | 2.32 (1.7-3.3) | 2.08 (1.5-2.7) | 0.84 (0.80-0.87) | 0.81 (0.78-0.85) |
| BB 3C 3D FNH Z1 PN    | 1.22 ( 1742 / 1428 ) | 4.08 (3.4-5.4) | 2.28 (2.0-2.8) | 0.79 (0.76-0.82) | 0.76 (0.72-0.80) |
| BB 3C 3D FNH Z1 MOSPN | 1.27 ( 3248 / 2557 ) | 3.54 (2.2-4.4) | 2.08 (1.8-2.6) | 0.81 (0.79-0.84) | 0.79 (0.76-0.81) |
| <Fit4: different emission model> | | | | | |
| BL 2C 3D FNH Z1 MOS1  | 1.35 (  728 /  539 ) | 1.19 | 1.19 | | |
| BL 2C 3D FNH Z1 MOS2  | 1.23 (  684 /  558 ) | 1.50 | 0.93 | | |
| BL 2C 3D FNH Z1 MOS12 | 1.36 ( 1509 / 1113 ) | 1.33 | 1.06 | | |
| BL 2C 3D FNH Z1 PN    | 1.44 ( 2060 / 1432 ) | 2.34 | 1.13 | | |
| BL 2C 3D FNH Z1 MOSPN | 1.43 ( 3670 / 2561 ) | 1.43 | 1.05 | | |
| <Fit5: different emission code - APEC> | | | | | |
| BL 3C 3D FNH Z1 MOS1  | 0.96 (  529 /  554 ) | 1.99 (1.4-2.2) | 1.61 (1.3-2.3) | 0.92 (0.82-1.00) | 0.82 (0.79-0.85) |
| BL 3C 3D FNH Z1 MOS2  | 1.10 (  588 /  535 ) | 2.12 (1.1-3.9) | 2.05 (1.4-3.4) | 0.83 (0.80-0.88) | 0.84 (0.81-0.93) |
| BL 3C 3D FNH Z1 MOS12 | 1.10 ( 1217 / 1109 ) | 1.98 (1.5-3.6) | 1.76 (1.3-2.2) | 0.85 (0.82-0.90) | 0.83 (0.81-0.86) |
| BL 3C 3D FNH Z1 PN    | 1.18 ( 1689 / 1428 ) | 3.53 (2.1-5.8) | 1.80 (1.6-2.6) | 0.81 (0.78-0.83) | 0.82 (0.79-0.84) |
| BL 3C 3D FNH Z1 MOSPN | 1.16 ( 2969 / 2557 ) | 2.80 (2.0-4.2) | 1.87 (1.6-2.1) | 0.82 (0.81-0.84) | 0.82 (0.81-0.84) |
| <Fit6, Fit7: without de-projection> | | | | | |
| BL 3C 2D FNH Z1 MOS1  | 1.04 (  577 /  554 ) | 1.93 (1.6-2.5) | 1.34 (1.2-1.7) | 0.83 (0.79-0.86) | 0.78 (0.71-0.83) |
| BL 3C 2D FNH Z1 MOS2  | 1.16 (  619 /  535 ) | 1.94 (1.6-2.3) | 1.49 (1.3-1.8) | 0.82 (0.79-0.85) | 0.80 (0.75-0.85) |
| BL 3C 2D FNH Z1 MOS12 | 1.17 ( 1293 / 1109 ) | 1.92 (1.6-2.3) | 1.45 (1.3-1.7) | 0.83 (0.81-0.85) | 0.79 (0.76-0.83) |
| BL 3C 2D FNH Z1 PN    | 1.21 ( 1729 / 1428 ) | 2.32 (2.1-2.7) | 1.56 (1.5-1.7) | 0.78 (0.76-0.80) | 0.74 (0.71-0.78) |
| BL 3C 2D FNH Z1 MOSPN | 1.21 ( 3094 / 2557 ) | 2.13 (1.9-2.3) | 1.49 (1.5-1.5) | 0.80 (0.79-0.82) | 0.77 (0.74-0.79) |
| | | | | | |
| BL 2C 2D FNH Z1 MOS1  | 1.33 (  741 /  558 ) | 1.10 | 0.74 | | |
| BL 2C 2D FNH Z1 MOS2  | 1.43 (  769 /  539 ) | 1.03 | 0.81 | | |
| BL 2C 2D FNH Z1 MOS12 | 1.44 ( 1604 / 1113 ) | 1.06 | 0.78 | | |
| BL 2C 2D FNH Z1 PN    | 1.55 ( 2218 / 1432 ) | 1.12 | 0.82 | | |
| BL 2C 2D FNH Z1 MOSPN | 1.51 ( 3880 / 2561 ) | 1.09 | 0.80 | | |
| <Fit8: varying N(H)> | | | | | |
| BL 3C 3D VNH Z1 MOS1  | 1.03 (  569 /  553 ) | 2.34 (1.4-3.8) | 1.37 (0.9-1.8) | | |
| BL 3C 3D VNH Z1 MOS2  | 1.14 (  610 /  534 ) | 2.08 (1.4-3.3) | 1.37 (1.2-2.1) | | |
| BL 3C 3D VNH Z1 MOS12 | 1.15 ( 1275 / 1108 ) | 1.98 (1.5-2.4) | 1.49 (1.2-1.7) | | |
| BL 3C 3D VNH Z1 PN    | 1.19 ( 1690 / 1427 ) | 2.47 (2.3-3.0) | 1.51 (1.4-1.7) | | |
| BL 3C 3D VNH Z1 MOSPN | 1.19 ( 3037 / 2556 ) | 2.03 (1.9-2.4) | 1.47 (1.4-1.6) | | |
| <Fit9-12: differrent groups of elements to vary> | | | | | |
| BL 3C 3D FNH Z2 MOS1  | 1.05 (  578 /  550 ) | 2.87 (1.8-3.7) | 1.75 (1.4-2.4) | | |
| BL 3C 3D FNH Z2 MOS2  | 1.17 (  619 /  531 ) | 2.39 (1.6-4.4) | 2.20 (1.5-4.6) | | |
| BL 3C 3D FNH Z2 MOS12 | 1.17 ( 1294 / 1105 ) | 2.29 (1.6-3.5) | 2.05 (1.5-2.8) | | |
| BL 3C 3D FNH Z2 PN    | 1.20 ( 1712 / 1424 ) | 3.67 (3.3-4.8) | 2.41 (2.1-2.8) | | |
| BL 3C 3D FNH Z2 MOSPN | 1.21 ( 3085 / 2553 ) | 3.23 (2.1-4.2) | 2.16 (1.8-2.7) | | |
| | | | | | |
| BL 3C 3D FNH Z3 MOS1  | 1.04 (  570 /  550 ) | 2.19 (1.6-4.2) | 1.68 (1.1-2.6) | | |
| BL 3C 3D FNH Z3 MOS2  | 1.12 (  596 /  531 ) | 1.76 (1.2-2.5) | 1.90 (1.2-3.3) | | |
| BL 3C 3D FNH Z3 MOS12 | 1.15 ( 1267 / 1105 ) | 1.99 (1.5-2.9) | 1.97 (1.4-2.4) | | |
| BL 3C 3D FNH Z3 PN    | 1.17 ( 1665 / 1424 ) | 2.69 (1.8-3.5) | 2.00 (1.5-2.3) | | |
| BL 3C 3D FNH Z3 MOSPN | 1.18 ( 3005 / 2553 ) | 2.49 (2.1-3.0) | 1.83 (1.6-2.5) | | |
| | | | | | |
| BL 3C 3D FNH Z4 MOS1  | 0.97   531   546 | [3.0] | [2.0] | | |
| BL 3C 3D FNH Z4 MOS2  | 1.05   550   527 | [3.0] | [2.0] | | |
| BL 3C 3D FNH Z4 MOS12 | 1.08  1192  1101 | [3.0] | [2.0] | | |
| BL 3C 3D FNH Z4 PN    | 1.17  1656  1420 | [3.0] | [2.0] | | |
| BL 3C 3D FNH Z4 MOSPN | 1.15  2930  2549 | [3.0] | [2.0] | | |
| | | | | | |
| BL 3C 3D FNH Z5 MOS1  | 0.97   521   538 | 2.14 (1.3-3.7) | 1.35 (1.0-2.0) | | |
| BL 3C 3D FNH Z5 MOS2  | 1.04   537   519 | 1.33 (1.0-1.9) | 1.53 (1.1-2.3) | | |
| BL 3C 3D FNH Z5 MOS12 | 1.07  1170  1093 | 1.55 (1.2-2.2) | 1.50 (1.2-1.7) | | |
| BL 3C 3D FNH Z5 PN    | 1.16  1639  1412 | 2.24 (2.0-2.4) | 1.89 (1.5-2.5) | | |
| BL 3C 3D FNH Z5 MOSPN | 1.15  2912  2541 | 1.91 (1.5-2.2) | 1.67 (1.4-2.1) | | |
| <Fit13: 5-shells> | | | | | |
| BL 3C 3D FNH Z1 MOS1  | 1.22   881   724 | 2.83 (1.7-4.3) | 1.75 (1.4-3.0) | 0.85 (0.82-0.96) | 0.80 (0.72-0.85) |
| BL 3C 3D FNH Z1 MOS2  | 1.24   873   702 | 2.31 (1.4-5.2) | 2.34 (1.6-3.1) | 0.83 (0.78-0.87) | 0.82 (0.79-0.86) |
| BL 3C 3D FNH Z1 MOS12 | 1.28  1852  1451 | 2.04 (1.7-3.7) | 2.18 (1.6-3.0) | 0.85 (0.83-0.87) | 0.81 (0.79-0.84) |
| BL 3C 3D FNH Z1 PN    | 1.22  2249  1846 | 4.42 (2.5-9.5) | 2.30 (1.8-3.2) | 0.79 (0.76-0.82) | 0.76 (0.71-0.80) |
| BL 3C 3D FNH Z1 MOSPN | 1.27  4220  3322 | 3.27 (2.3-4.5) | 2.08 (1.9-2.7) | 0.81 (0.79-0.83) | 0.79 (0.76-0.81) |
| <Fit14: 6-shells> | | | | | |
| BL 3C 3D FNH Z1 MOS1  | 1.38  1262   946 | 2.36 (1.4-4.1) | 1.92 (1.5-3.1) | 0.86 (0.80-0.90) | 0.80 (0.74-0.84) |
| BL 3C 3D FNH Z1 MOS2  | 1.17  1049   895 | 2.35 (1.3-3.7) | 2.44 (1.7-3.6) | 0.82 (0.81-0.87) | 0.78 (0.76-0.86) |
| BL 3C 3D FNH Z1 MOS12 | 1.34  2473  1871 | 2.68 (1.9-3.9) | 1.74 (1.5-2.7) | 0.84 (0.80-0.86) | 0.82 (0.78-0.85) |
| BL 3C 3D FNH Z1 PN    | 1.20  2809  2350 | 4.49 (2.9-7.3) | 2.31 (1.9-2.9) | 0.79 (0.77-0.83) | 0.76 (0.70-0.80) |
| BL 3C 3D FNH Z1 MOSPN | 1.29  5439  4221 | 2.90 (2.3-4.6) | 2.15 (1.9-2.7) | 0.82 (0.80-0.84) | 0.78 (0.76-0.81) |

Note.

- The error in the parenthesis is in the 90% confidence level.
- For those with poor spectral fitting (reduce Chi^2 > 1.3),
  errors are not meaningful and not listed, except FIT3 and 13-14 for
  comparison (see section 3.3).
- kT is the temperature of the soft thermal component.

- Codes used in the "method" column:

1st column - background spctra
   BL: background spectra are taken from the local region (see Figure 2)
   BE: background spectra are taken from the edge of the field of view (see Figure 2), then scaled by arf
   BB: background spectra are taken from the blank field, then scaled by counts in the BKL region

2nd column - emission models
   2C: 2-component model, 1 soft thermal (vmekal) + 1 hard (7 keV bremsstrahlung)
   3C: 3-component model, 2 soft thermal (vmekal + 1.4 keV vmekal) + 1 hard (7 keV bremsstrahlung)

3rd column - projection
   2D: no de-projection
   3D: projection of 3-dimensional models and compare with data

4th column - NH
   FHN: N(H) is fixing at the galactic value ($5 \times 10^{20}$ cm$^{-2}$)
   VNH: N(H) is free to vary

5th column - grouping elements
   Z1: all heavy elements vary together
   Z2: Si and S vary together and the other elements vary with Fe
   Z3: elements lighter than Ca vary with Si and the rest elements vary with Fe
   Z4: Fe and Si are fixed. S, Mg and O vary independently and the rest elements vary as in Z3
   Z5: Fe, Si, S, Mg and O vary independently and the rest elements vary as in Z3

6th column - instrument
   MOS1, MOS2 and PN:  fitting individually
   MOS12: fitting jointly for MOS1 + MOS2
   MOSPN: fitting jointly for MOS1 + MOS2 + PN

Table 2. Radial variations of spectral parameters

| | r=0-1' (0-20 kpc) | r=1-2' (20-41 kpc) | r=2-3' (41-61 kpc) | r=3-5' (61-102 kpc) | r=5-7' (102-143 kpc) | r=7-10' (143-204 kpc) |
|---|---|---|---|---|---|---|
| **3-component models:** | | | | | | |
| **\<Z1 - Fit1\>** | | | | | | |
| T(1)  | 0.82 ( 0.79 -  0.84) | 0.78 ( 0.75 -  0.80) | 0.71 ( 0.65 -  0.78) | 0.63 ( 0.54 -  0.73) | | |
| Fx(1) | 5.25 ( 3.48 -  6.89) | 4.26 ( 3.47 -  5.33) | 0.83 ( 0.56 -  1.10) | 0.81 ( 0.50 -  0.98) | | |
| Fx(2) | 6.75 ( 4.86 -  8.71) | 12.62 (10.57 - 14.93) | 11.03 ( 9.09 - 12.12) | 13.08 (11.63 - 14.85) | | |
| Fx(3) | 0.82 ( 0.07 -  1.49) | 1.43 ( 0.83 -  2.05) | 0.91 (   -   ) | 2.77 ( 1.66 -  3.64) | | |
| Fe    | 2.85 ( 2.23 -  3.79) | 2.16 ( 1.80 -  2.58) | 1.03 ( 0.92 -  1.27) | 0.71 ( 0.62 -  0.82) | | |
| **\<Z2 - Fit8\>** | | | | | | |
| T(1)  | 0.81 ( 0.79 -  0.84) | 0.78 ( 0.74 -  0.81) | 0.71 ( 0.60 -  0.81) | 0.64 ( 0.54 -  0.73) | | |
| Fx(1) | 5.18 ( 3.84 -  7.95) | 4.16 ( 3.31 -  5.22) | 0.79 ( 0.50 -  1.08) | 0.82 ( 0.50 -  1.16) | | |
| Fx(2) | 6.75 ( 5.07 -  9.78) | 12.60 (10.41 - 15.08) | 11.01 ( 8.67 - 11.99) | 12.96 (11.63 - 14.85) | | |
| Fx(3) | 0.90 ( 0.15 -  1.62) | 1.57 ( 0.92 -  2.19) | 0.97 (   -   ) | 2.94 ( 1.90 -  3.87) | | |
| Fe    | 3.23 ( 2.15 -  4.21) | 2.16 ( 1.82 -  2.65) | 1.00 ( 0.91 -  1.25) | 0.72 ( 0.62 -  0.83) | | |
| Si    | 3.30 ( 2.75 -  3.83) | 1.94 ( 1.63 -  2.28) | 0.89 ( 0.70 -  1.16) | 0.70 ( 0.54 -  0.86) | | |
| **\<Z3 - Fit9\>** | | | | | | |
| T(1)  | 0.81 ( 0.78 -  0.84) | 0.77 ( 0.72 -  0.80) | 0.71 ( 0.56 -  0.82) | 0.59 ( 0.51 -  0.76) | | |
| Fx(1) | 4.76 ( 3.37 -  6.68) | 3.78 ( 2.93 -  4.64) | 0.72 ( 0.40 -  1.03) | 0.77 ( 0.47 -  1.13) | | |
| Fx(2) | 7.07 ( 4.90 -  9.67) | 12.84 (10.17 - 15.19) | 10.90 ( 8.52 - 12.21) | 12.55 (11.64 - 15.59) | | |
| Fx(3) | 1.03 ( 0.19 -  1.73) | 1.77 ( 1.05 -  2.57) | 1.21 (   -   ) | 3.56 ( 1.80 -  4.41) | | |
| Fe    | 2.49 ( 1.83 -  3.53) | 1.83 ( 1.55 -  2.31) | 0.97 ( 0.86 -  1.29) | 0.77 ( 0.61 -  0.84) | | |
| Si    | 1.90 ( 1.53 -  2.62) | 1.32 ( 1.03 -  1.80) | 0.78 ( 0.60 -  1.14) | 0.65 ( 0.44 -  0.77) | | |
| **\<Z4 - Fit10\>** | | | | | | |
| T(1)  | 0.78 ( 0.75 -  0.81) | 0.74 ( 0.71 -  0.77) | 0.71 ( 0.62 -  0.76) | 0.57 ( 0.48 -  0.69) | | |
| Fx(1) | 4.73 ( 4.32 -  5.19) | 3.75 ( 3.37 -  4.12) | 0.72 ( 0.46 -  0.99) | 0.70 ( 0.46 -  1.01) | | |
| Fx(2) | 7.10 ( 6.07 -  7.97) | 12.88 (12.05 - 13.83) | 10.99 (10.32 - 13.58) | 12.94 (12.25 - 13.58) | | |
| Fx(3) | 1.06 ( 0.41 -  1.73) | 1.75 ( 1.07 -  2.34) | 1.09 (   -  8.39) | 3.06 ( 2.10 -  3.93) | | |
| S     | 2.16 ( 1.43 -  2.95) | 1.23 ( 0.84 -  1.60) | 0.76 ( 0.45 -  1.07) | 0.44 ( 0.22 -  0.66) | | |
| Mg    | 3.10 ( 2.40 -  3.85) | 1.63 ( 1.19 -  1.84) | 0.56 ( 0.15 -  1.00) | 0.59 ( 0.28 -  0.90) | | |
| O     | 1.49 ( 1.02 -  1.99) | 1.13 ( 0.83 -  1.41) | 0.75 ( 0.49 -  1.06) | 0.58 ( 0.37 -  0.81) | | |
| **\<Z5 - Fit11\>** | | | | | | |
| T(1)  | 0.79 ( 0.76 -  0.82) | 0.74 ( 0.71 -  0.78) | 0.71 ( 0.58 -  0.78) | 0.57 ( 0.50 -  0.72) | | |
| Fx(1) | 4.80 ( 3.35 -  6.37) | 3.68 ( 3.00 -  4.66) | 0.70 ( 0.38 -  1.01) | 0.62 ( 0.40 -  0.99) | | |
| Fx(2) | 7.25 ( 5.33 -  9.47) | 13.09 (10.55 - 15.31) | 10.96 ( 9.22 - 13.29) | 13.16 (10.50 - 14.30) | | |
| Fx(3) | 0.75 (   -  1.78) | 1.55 ( 0.74 -  2.43) | 1.16 (   -  4.56) | 2.85 ( 1.87 -  4.26) | | |
| Fe    | 1.91 ( 1.46 -  2.24) | 1.66 ( 1.43 -  2.06) | 1.01 ( 0.82 -  1.09) | 0.67 ( 0.60 -  0.81) | | |
| Si    | 2.05 ( 1.59 -  2.48) | 1.55 ( 1.29 -  1.94) | 0.94 ( 0.62 -  1.20) | 0.61 ( 0.47 -  0.89) | | |
| S     | 1.23 ( 0.62 -  1.67) | 0.97 ( 0.63 -  1.39) | 0.79 ( 0.40 -  1.10) | 0.41 ( 0.20 -  0.71) | | |
| Mg    | 1.93 ( 1.35 -  2.35) | 1.25 ( 0.88 -  1.76) | 0.59 ( 0.08 -  1.01) | 0.52 ( 0.24 -  0.98) | | |
| O     | 0.68 ( 0.25 -  0.94) | 0.78 ( 0.52 -  1.16) | 0.79 ( 0.29 -  0.86) | 0.51 ( 0.36 -  0.97) | | |
| **Two-component models:** | | | | | | |
| **\<3D - Fit4\>** | | | | | | |
| T(1+2) | 0.99 ( 0.97 -  1.01) | 1.03 ( 1.02 -  1.04) | 1.41 ( 1.38 -  1.43) | 1.41 ( 1.38 -  1.43) | | |
| **\<2D - Fit6\>** | | | | | | |
| T(1+2) | 1.03 ( 1.03 -  1.04) | 1.10 ( 1.09 -  1.11) | 1.37 ( 1.35 -  1.39) | 1.40 ( 1.38 -  1.43) | | |
| **Extra shells:** | | | | | | |
| **\<5 shells - Fit13\>** | | | | | | |
| T(1)  | 0.81 ( 0.79 -  0.83) | 0.79 ( 0.76 -  0.81) | 0.71 ( 0.65 -  0.77) | 0.66 ( 0.53 -  0.76) | 0.71 ( 0.70 -  0.73) | |
| Fx(1) | 5.29 ( 3.70 -  8.53) | 4.55 ( 3.40 -  5.36) | 1.15 ( 0.87 -  1.44) | 1.93 ( 0.93 -  3.06) | 2.46 ( 2.23 -  2.70) | |
| Fx(2) | 7.00 ( 5.14 -  9.80) | 13.37 (10.62 - 15.14) | 12.44 (10.96 - 13.99) | 18.19 (16.29 - 19.54) | 11.99 (11.03 - 12.93) | |
| Fx(3) | 0.81 ( 0.52 -  1.38) | 1.25 ( 0.64 -  1.91) | 0.74 ( 0.13 -  1.30) | 1.02 (   -   ) | 1.67 ( 1.06 -  2.28) | |
| Fe    | 3.27 ( 2.29 -  4.45) | 2.08 ( 1.94 -  2.65) | 1.08 ( 0.94 -  1.24) | 0.71 ( 0.64 -  0.81) | 0.75 ( 0.69 -  0.83) | |
| **\<6 shells - Fit14\>** | | | | | | |
| T(1)  | 0.82 ( 0.80 -  0.84) | 0.78 ( 0.76 -  0.81) | 0.71 ( 0.65 -  0.76) | 0.69 ( 0.59 -  0.78) | 0.10 ( 0.49 -  0.88) | 0.71 ( 0.70 -  0.74) |
| Fx(1) | 5.32 ( 3.67 -  6.97) | 4.51 ( 3.72 -  5.31) | 1.14 ( 0.83 -  1.45) | 1.85 ( 1.39 -  2.34) | 2.00 (   -   ) | 4.35 ( 3.89 -  4.89) |
| Fx(2) | 7.01 ( 5.75 -  8.52) | 13.40 (11.05 - 14.63) | 12.47 (10.02 - 13.99) | 18.37 (16.29 - 19.86) | 12.86 (11.41 - 14.49) | 13.34 (11.75 - 14.49) |
| Fx(3) | 0.77 ( 0.14 -  1.39) | 1.27 ( 0.70 -  1.93) | 0.72 ( 0.12 -  1.33) | 0.86 (   -  4.84) | 1.14 ( 0.29 -  1.77) | 0.00 (   -   ) |
| Fe    | 2.90 ( 2.27 -  4.63) | 2.15 ( 1.88 -  2.75) | 1.07 ( 0.95 -  1.24) | 0.77 ( 0.69 -  0.88) | 0.41 ( 0.34 -  0.50) | 1.11 ( 0.97 -  1.27) |

Note.

T(1) : temperature of the soft thermal component  
Fx(1): X-ray flux of the soft thermal component in unit of $10^{-13}$ erg sec$^{-1}$ cm$^{-2}$ at 0.3-8 keV  
Fx(2): X-ray flux of the 1.4 keV thermal component in unit of $10^{-13}$ erg sec$^{-1}$ cm$^{-2}$ at 0.3-8 keV  
Fx(3): X-ray flux of the 7 keV LMXB component in unit of $10^{-13}$ erg sec$^{-1}$ cm$^{-2}$ at 0.3-8 keV  
T(1+2): average temperature of the soft + 1.4 keV components, determined in a 2-component model

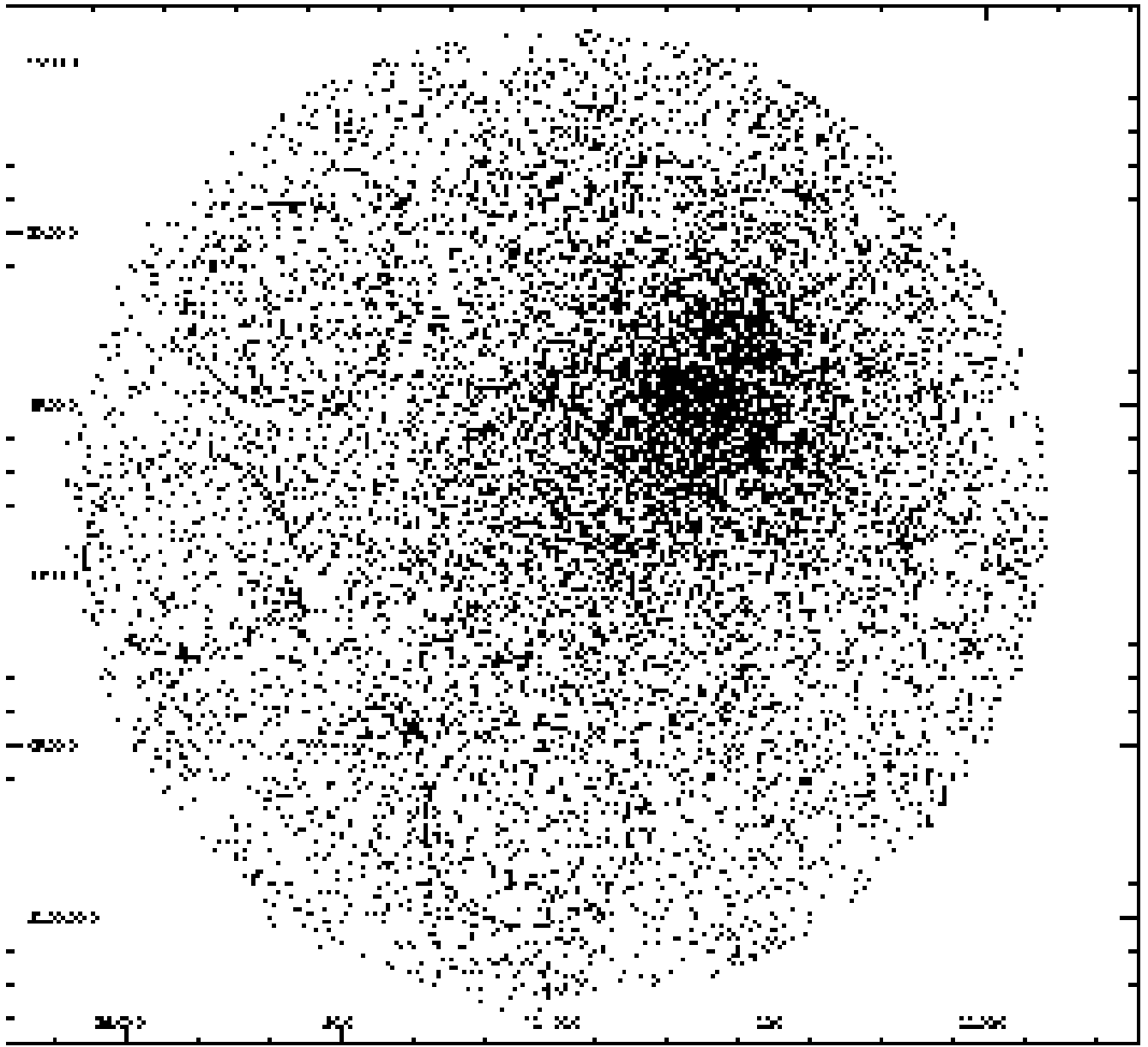

Figure 1:

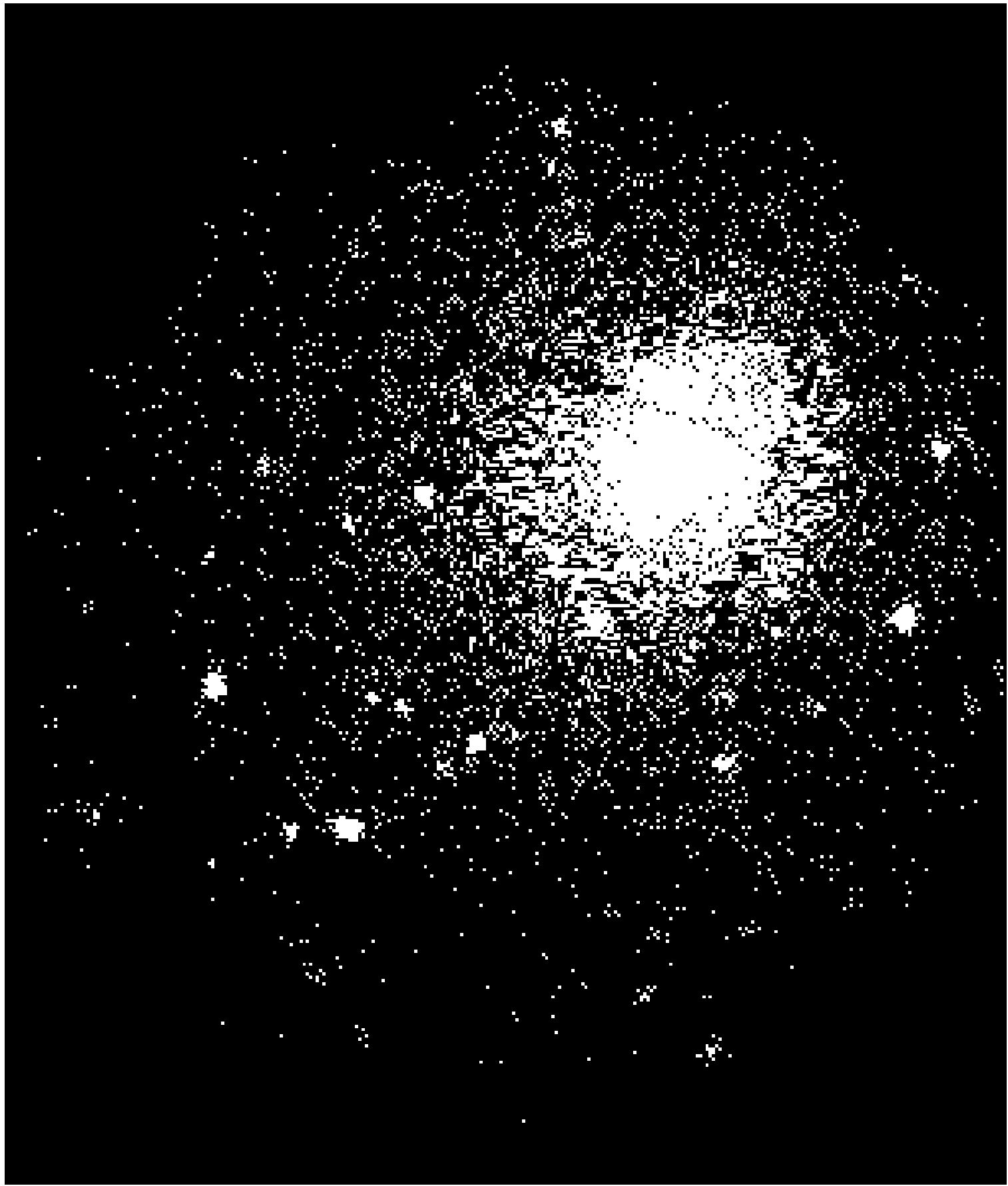

Figure 2:

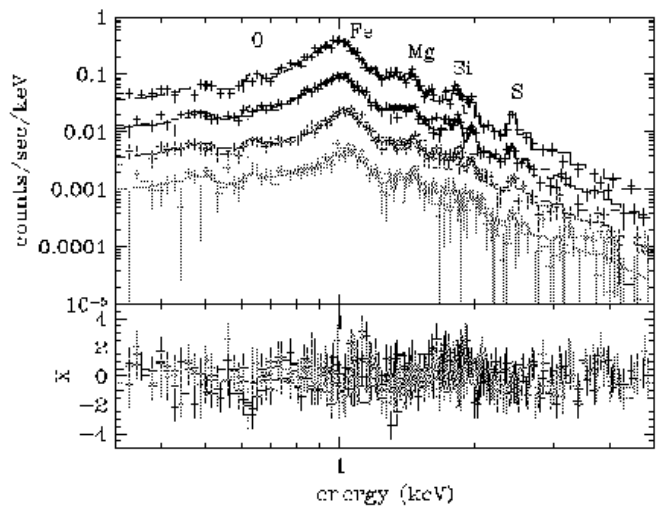

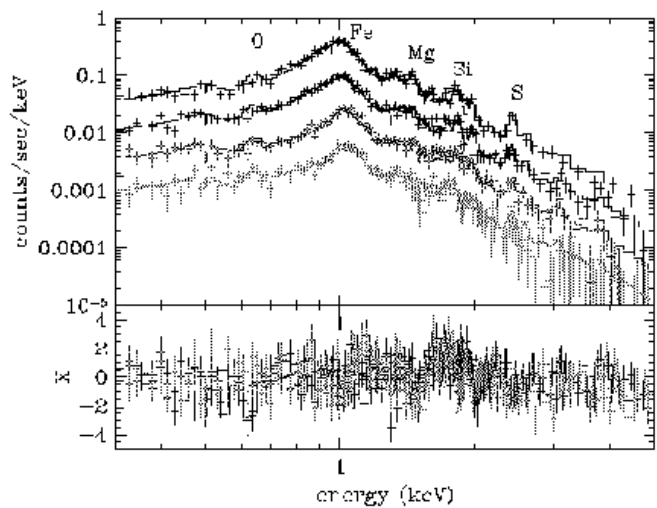

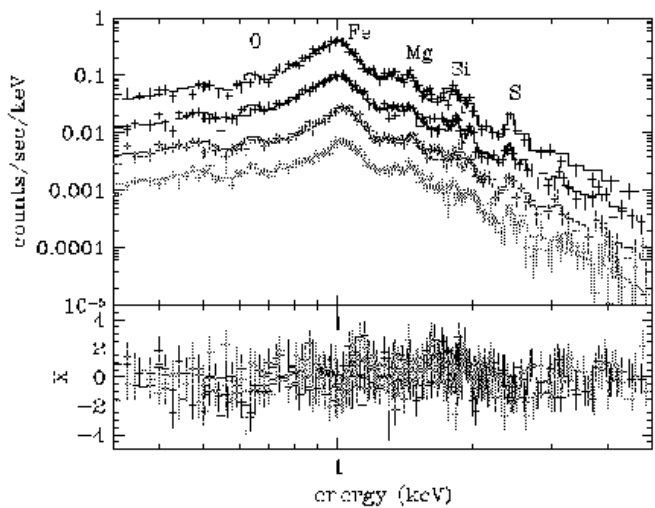

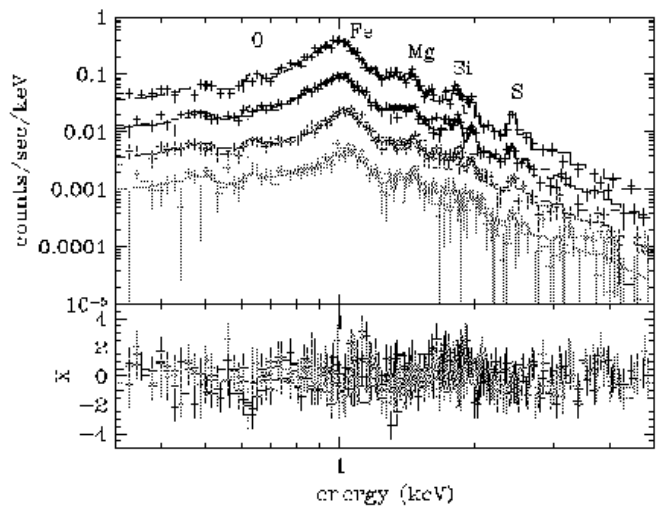
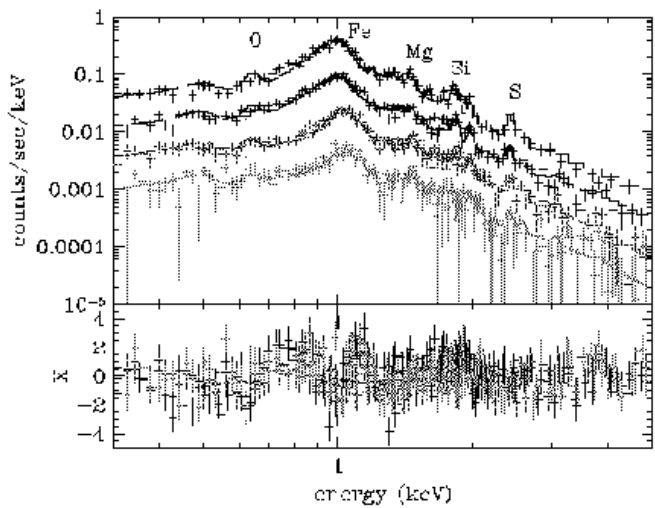
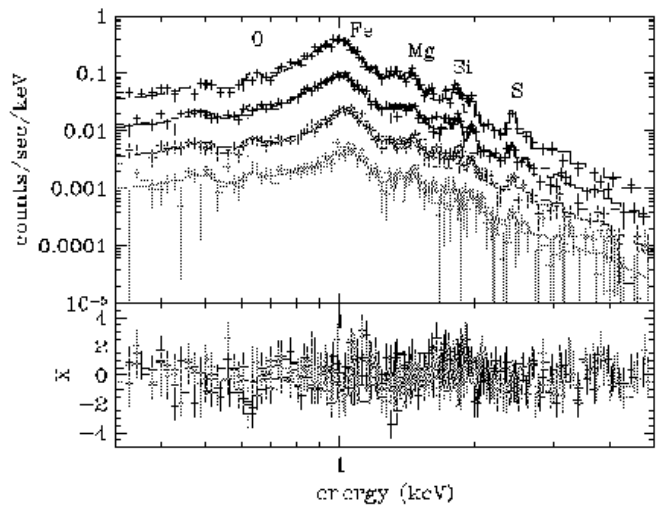
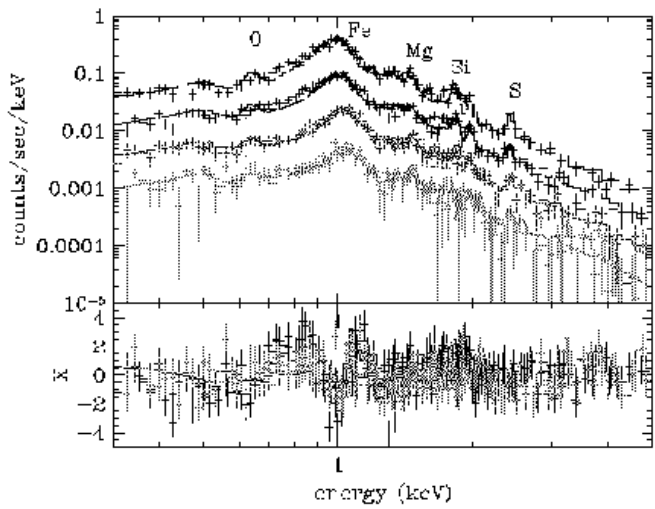

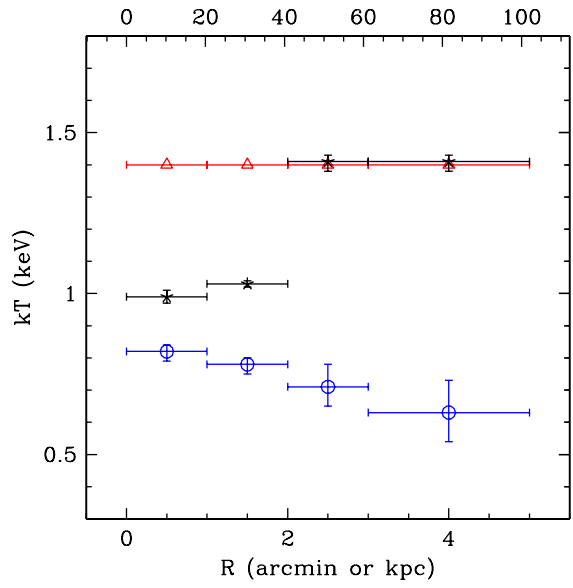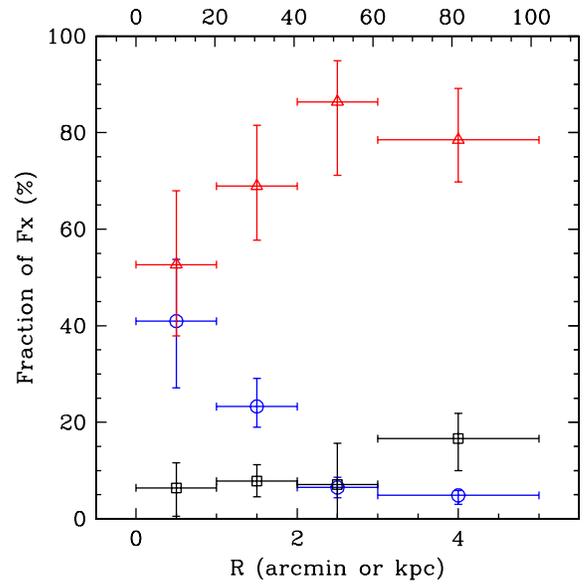

Figure 5:

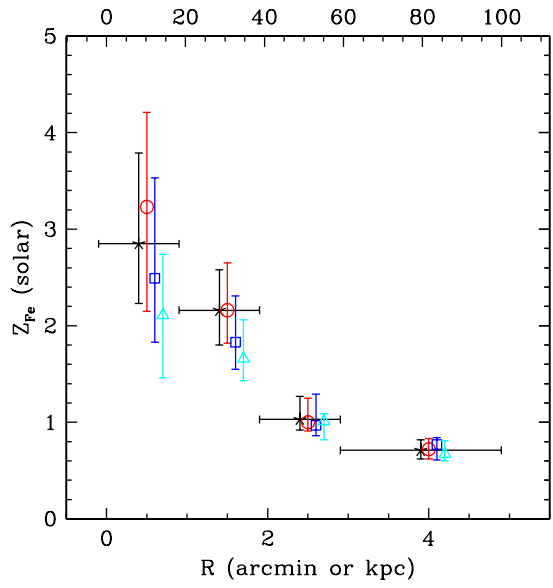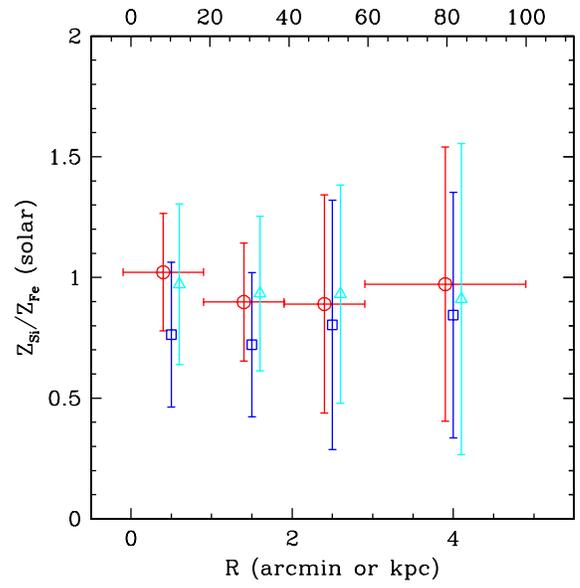

Figure 6: